\documentclass[aps,prd]{revtex4}
\usepackage{graphicx}

\begin{document}

\title{ Features in the Primordial Spectrum from WMAP: A Wavelet Analysis}
\author{Arman Shafieloo, Tarun Souradeep}
\affiliation{Inter-University Centre for Astronomy and Astrophysics (IUCAA), 
Ganeshkhind, Pune-411007, India}
\author{P. Manimaran, Prasanta K. Panigrahi, Raghavan Rangarajan}
\affiliation{Physical Research Laboratory, Ahmedabad 380 009, India}

\date{November 2006}


\def\be{\begin{equation}}
  \def\ee{\end{equation}}
\def\bea{\begin{eqnarray}}
  \def\eea{\end{eqnarray}}
\def\ie{{\it i.e.}\ }
\def\eg{{\it e.g.}\ }

\begin{abstract}
 Precise measurements of the anisotropies in the cosmic microwave
background enable us to do an accurate study on the form of the
primordial power spectrum for a given set of cosmological
parameters. In a previous paper~\cite{prd04}, we implemented an
improved (error sensitive) Richardson-Lucy deconvolution algorithm on
the measured angular power spectrum from the first year of WMAP data
to determine the primordial power spectrum assuming a concordance
cosmological model. This recovered spectrum has a likelihood far
better than a scale invariant, or, `best fit' scale free spectra
($\Delta\ln{\cal L}\approx 25$ {\it w.r.t.} Harrison Zeldovich, and,
$\Delta\ln{\cal L}\approx 11$ {\it w.r.t.} power law with
$n_s=0.95$). In this paper we use Discrete Wavelet Transform
(DWT) to decompose the local features of the recovered spectrum individually
to study their effect and significance on the recovered angular power
spectrum  and hence the likelihood. We show that besides the infra-red
cut off at the horizon scale, the associated features of the primordial power spectrum
around the horizon have a significant effect on improving the
likelihood. The strong features are localized at the horizon scale.
\end{abstract}

\maketitle


\section{Introduction}
\label{sec:intro}
Observational comparison of cosmological models based on structure
formation in the universe necessarily depends on the assumed initial
conditions describing the primordial seed perturbations. Inflation,
besides resolving a number of problems of classical Big Bang theory,
provides us a mechanism for generating these correlated primordial
perturbations~\cite{inflation1,inflation2,inflation3}. Precision
measurements of anisotropies in the cosmic microwave background, and
also of the clustering of large scale structure, suggest that the
primordial density perturbation is dominantly adiabatic and has a
nearly scale invariant spectrum~\cite{sel04,sper_wmap06}. This is in
good agreement with most simple inflationary scenarios which predict
power law or scale invariant forms of the primordial
perturbation. The data have also been used widely to put constraints on different parametric forms of primordial spectrum, mostly motivated by inflation~\cite{hiranya06,covi06,bridges06,jerome06}.  

However, despite the strong theoretical appeal and simplicity of a
featureless primordial spectrum, it is important to determine the
shape of the primordial power spectrum directly from observations with
minimal theoretical bias. Many model independent searches have been
made to look for features in the CMB primordial power
spectrum~\cite{bridle03,hanne04,pia03,pia05,sam06}. Accurate
measurements of the angular power spectrum over a wide range of
multipoles by WMAP has opened up the possibility of deconvolving the
primordial power spectrum for a given set of cosmological
parameters~\cite{max_zal02,mat_sas0203,prd04,bump05,armarun06,kog03}. Theoretical
motivations and models that give features in the power spectrum have
also been studied and compared in post-WMAP literature~\cite{contaldi03,manoj03,takahashi,sarkar04,sarkar05,sin_sour06,cline06,amjad06}. The primordial power
spectrum has been deconvolved directly from the angular power spectrum
of CMB anisotropy measured by WMAP using an improved implementation of
the Richardson-Lucy algorithm~\cite{prd04}. The most prominent
feature of the recovered spectrum is a sharp infra-red cut-off on the
horizon scale. It also has a localized excess just above the cut-off
which leads to great improvement of likelihood over the simple
monotonic forms of model infra-red cut-off spectra considered widely
in the post-WMAP-1 literature. Interestingly, similar features were
also detected by method of regularized least squares~\cite{bump05}. The significant improvement in the likelihood clearly
shows the importance of features in the primordial power spectrum. 

{\em The goal of the present paper is to demonstrate the application of wavelets in identifying the statistically significant features in a deconvolved power spectrum.}
We use Discrete Wavelet Transform (henceforth DWT) to
identify features in the recovered primordial power spectrum at
different resolutions and at different locations in $k$ space.  Starting from the coarsest primordial power spectrum we systematically add
variations on different resolutions and obtain the angular power spectrum.
We then compute the likelihood of the reconstructed primordial power
spectrum by comparing the angular power spectrum with the WMAP data.
The improvement in the likelihood allows us to quantify the
significance of different features.

\section{Richardson-Lucy deconvolution Method}

The Richardson-Lucy (RL) algorithm was developed and is widely used in
the context of image reconstruction in
astronomy~\cite{lucy74,rich72}. However, the method has also been
successfully used in cosmology, to deproject the $3$-D correlation
function and power spectrum from the measured $2$-D angular
correlation and $2$-D power spectrum~\cite{baug_efs93,baug_efs94}.

The angular power spectrum, $C_l$, is a convolution of the initial power spectrum $P(k)$ generated in the early universe with a radiative transport kernel, $G(l,k)$, that is determined by the values of the cosmological parameters. In our application, we solve the inverse problem of determining
the primordial power spectrum, $P(k)$, from the measured angular power
spectrum, $C_l$, using the relation

\be C_l\,= \sum_i G(l,k_i )\,P(k_i). 
\label{clsum}
\ee 

In the above equation, the {\em `target'} measured angular power
spectrum, $C_l\equiv C_l^D$, is the data given by observations, and
the radiative transport kernel,
\be G(l,k_i) = \frac{\Delta
k_i}{k_i}\,|{\Delta_{Tl}(k_i,\eta_0)}|^2\,,
\label{glk}
\ee encodes the response of the present multipoles of the CMB
perturbed photon distribution function $\Delta_{Tl}(k_i,\eta_0)$ to
unit of power per logarithm interval of wavenumber, $k$, in the
primordial perturbation spectrum. The kernel $G(l,k)$ is completely
fixed by the cosmological parameters of the {\em `base'} cosmological
model. 
Obtaining $P(k)$ from the measured $C_l$, for a given
$G(l,k)$, is clearly a deconvolution problem.  An important feature of
the problem is that $C_l^D$, $G(l,k)$ and $P(k)$ are all positive
definite.

 In Ref.~\cite{prd04} we (A.S. and T.S.) employ an improved RL method 
to solve the inverse problem for 
$P(k)$ in Eq.~(\ref{clsum}).  The advantage
of RL method is that positivity of the recovered $P(k)$ is
automatically ensured, given $G(l,k)$ is positive definite and $C_l$'s
are positive. The RL method is readily derived from elementary
probability theory on distributions~\cite{lucy74} and is an iterative
method which can be neatly encoded into a simple recurrence
relation. The power spectrum $P^{(i+1)}(k)$ recovered after iteration
$(i+1)$ is given by

\be
P^{(i+1)}(k)- P^{(i)}(k)\,=\,P^{(i)}(k)\,\sum_l\,G(l,k)\,\frac
{C^D_l\,-C_l^{(i)}}{C_l^{(i)}}
\label{RLstd}
\ee 
where $C^D_l$ is the measured data (target) and $C_l^{(i)}$ is the
angular power spectrum at the $i^{\rm th}$ iteration obtained from
Eq.~(\ref{clsum}) using the recovered power spectrum $P^{(i)}(k)$.
Eq.~(\ref{RLstd}), together with Eq.~(\ref{clsum}) for obtaining $C_l^{(i)}$
from $P^{(i)}(k)$ completely summarizes the standard RL method.
The final recovered power spectrum is independent of the initially chosen
$P^{(1)}(k)$ for reasonable forms of $P^{(1)}(k)$ (see discussion in the
Appendix of Ref.~\cite{prd04}).

Due to noise and sample variance, the data $C_l^D$ is measured within
some non-zero error bars $\sigma_l$. The standard RL method does not
incorporate the error information at all and tends to iterate to fit
features of the noise, as well.  In our problem, the problem manifests
itself as a non-smooth deconvolved spectrum $P(k)$ from the binned
data that has poor likelihood with the full WMAP spectrum data.  
In Ref.~\cite{prd04} we devise a novel procedure  to make the RL method sensitive to the errors $\sigma_l$ by modifying Eq.~(\ref{RLstd}) to

\be P^{(i+1)}(k)- P^{(i)}(k)\,=\,P^{(i)}(k)\,\sum_l\,G(l,k)\,\frac
{C_l^D\,-C_l^{(i)}}{C_l^{(i)}}\,\,\,\tanh^2\,
\left[\frac{(C^D_l\,-C_l^{(i)})^2}{{\sigma_l}^2}\right].
\label{RLerr}
 \ee The idea is to employ a `convergence' function to progressively
weigh down the contribution to the correction $P^{(i+1)}-P^{(i)}$ from
a multipole bin where $C_l^{(i)}$ is close to $C_l^D$ within the error
bar $\sigma_l$. This innovation significantly improves the WMAP
likelihood of the deconvolved spectrum. The final form of the
recovered spectrum was obtained after smoothing the spectrum by a
simple ``bowler-hat'' smoothing kernel, and varying width of smoothing of
the kernel to get the best likelihood for the recovered angular power
spectrum~\cite{notearmarun06}.

In Ref.~\cite{prd04}, we have carefully verified our 
improved Richardson-Lucy method on some toy models before applying it to the 
real data. The method has been shown to work well after a careful analysis 
of artifacts and convergence issues.  Subsequently,
a very similar primordial power spectrum has been obtained in 
Ref.~\cite{bump05} by deconvolution of the angular power spectrum from 
WMAP data using the method of regularized least squares. 
It is indeed of interest to apply other deconvolution methods and to compare 
the results.

\begin{figure}[h] 
  \includegraphics[scale=0.4, angle=0]{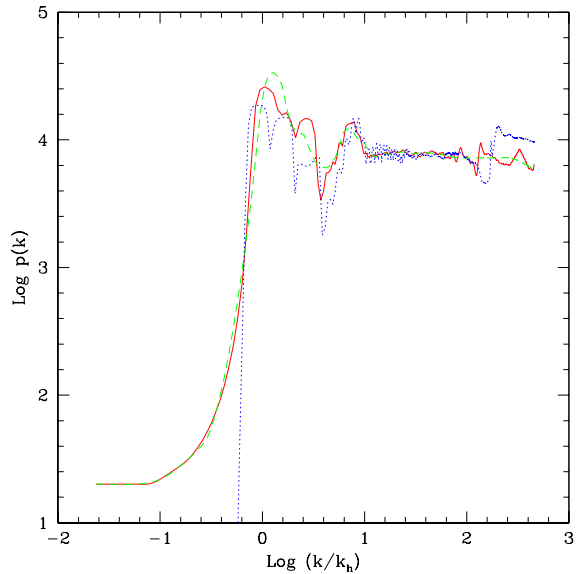}
  \caption{ The recovered spectrum for a flat cosmological model with
  $\tau$=0.0, $h$=0.71, $\Omega_b h^2$=0.0224 and
  $\Omega_{\Lambda}$=0.73 (red-solid line). The green-dashed line shows the
  reconstructed spectrum, using DWT, incorporating the coarse
  behavior and the low frequency features captured by the highest
  level wavelet coefficients.  Even at this stage, the likelihood for
  the recovered angular power spectrum ( $\chi^{2}_{\rm eff}=974.1$)
  is better than the likelihood for the best fit power-law spectrum (
  $\chi^{2}_{\rm eff}=978.6$). The blue-dotted line is the recovered spectrum using WMAP-3 data for the same background cosmology. This recovered spectrum gives $\Delta \chi^2_{eff} =-15.93$ with respect to the best power law primordial spectrum in the whole cosmological parameter space.} 
\label{main}
\end{figure}

\section{Discrete Wavelet Transform}

Wavelet transforms provide a powerful tool for the analysis of
transient and non-stationary data and is particularly useful in
picking out characteristic local variations at different resolutions. This linear transform separates a data set in the form of
low-pass or average coefficients, which reflect the average behavior
of the data, and wavelet or high-pass coefficients at different
levels, which capture the variations at corresponding resolutions. As
compared to Fourier or window Fourier transform, wavelets allow
optimal ``time-frequency'' localization simultaneously in the real, as well as,
Fourier domain. The vocabulary of DWT stems from applications in one
dimensional time-stream signal trains, but has found wide application
in signal in other domains and dimensions. Specifically in our case,
the `signal' being transformed is the power spectrum, $P(k)$, a one dimensional function of wavenumber, $k$.

Wavelets are an orthonormal basis of small waves, with their
variations primarily concentrated in a finite region, which makes them
ideal for analyzing localized `transient' signals. Wavelets can be
continuous or discrete. In the latter case, the basis elements are
strictly finite in size, enabling them to achieve localization, while
disentangling characteristic variations at different frequencies
\cite{Daubechies}. This is the primary reason for us to employ
discrete wavelets for our analysis.

In the construction of the basis set for discrete wavelet transform,
one starts with the scaling function $\varphi(x)$ (father wavelet) and
the mother wavelet $\psi(x)$, whose height and width are arbitrary:
$ \int \varphi \,dx=A ,~\int\psi\, dx=0,~ \int \varphi\,\psi \,dx=0, \int
|\varphi|^2 \,dx=1,~\int |\psi|^2\,dx=1$,  where $A$ is an arbitrary
constant.
In addition to the scaling and wavelet functions, their translates
$ \psi_{j,m}=2^{j/2}\,\psi(2^j x-m),
\,\varphi_{j,m}=2^{j/2}\varphi(2^j x-m),$
are also square integrable at different resolutions. Here, $m$ and $j$
respectively are the translation and scaling parameters, both
taking integral values with $-\infty\leq m \leq +\infty$.
We start with the resolution value $j =0$
and increase it by integral
units. The original mother wavelet corresponds to $\psi_{0,0}$,
and the father wavelet is given by $\varphi_{0,0}$. Higher values
of $j$ lead to the so called daughter wavelets which are of a
similar form as the mother wavelet, except that their width and
height differ by a factor of $2^{j}$ and $2^{j/2}$ respectively,
at successive levels. The translation unit $m/2^j$ is also
commensurate with the thinner size of the daughter wavelet at
resolution $j$. In the limit $j\rightarrow \infty$, these basis
functions form a complete orthonormal set.
It needs to be pointed out that for a discrete data set with a finite
number of points $N$ the maximum value of $j$ is the largest integer
less than or equal to $\log_2 N$.

 A signal $f(x)$ can then
be expanded as
\begin{eqnarray} f(x)=\sum_{m=-\infty}^{+\infty}c_{0,m}
\varphi_{0,m}(x)+\sum_{m=-\infty}^{+\infty}\sum_{j\geq
0}^{}d_{j,m} \psi_{j,m}(x)~~.\end{eqnarray}
Here, $c_{j,m}$'s are the low-pass coefficients and $d_{j,m}$'s are
the high-pass or wavelet coefficients, respectively capturing the
average part and variations of the signal at resolution $j$ and location
$m$.
In practice, for a finite data set
one starts with the highest level of resolution and
progressively moves to resolution on grosser scales keeping in mind the
physics of the problem and the size of the data set.  The lowest level of
resolution one chooses
then corresponds to $j=0$ and the higher levels correspond to larger
values of $j$.
For the discrete wavelets, the property of multi-resolution
analysis (MRA) leads to $c_{j,m}=\sum_{n} h(n-2m)c_{j+1,n}$ and
$d_{j,m}=\sum_{n} \tilde{h}(n-2m)c_{j+1,n},$ where $h(n)$ and
${\tilde{h}}(n)$ are respectively the low-pass (scaling function) and
high-pass (wavelet) filter coefficients, which differ for different
wavelets. Thus, both low-pass and high-pass coefficients at a resolution $j$ can be obtained from the low-pass coefficients at a higher resolution $j+1$. The low-pass coefficients $c_{j+1,m}$ are obtained though the
convolution of the signal $f(x)$ with the scaling function
$\varphi_{j+1,m}=2^{(j+1)/2}\varphi(2^{j+1} x-m)$. For a fixed $m$, in
the limit $j\rightarrow \infty$, the scaling function becomes a Dirac
delta function and hence the corresponding low-pass coefficient is the
signal itself at point $m$.  This implies that, starting from the
finest resolution of the signal, one can construct both scaling and
wavelet functions, by convolution with the corresponding filter
coefficients. Hence one can carry out the wavelet decomposition, as
also the inverse transform, with the help of $h(n)$ and
${\tilde{h}}(n)$, without explicitly knowing the basis set. In this
respect, wavelet transform is significantly different from the Fourier
transform.

For the Haar wavelet, $h(0)=h(1)=\frac{1}{\sqrt{2}}$ and
$\tilde{h}(0)=-\tilde{h}(1)=\frac{1}{\sqrt{2}}$. The Haar basis is
unique, since it is the only wavelet which is symmetric and
compactly supported. In a level one Haar wavelet decomposition,
the level-I low-pass (average) and high-pass (wavelet or detail)
coefficients are respectively given by the nearest neighbor
averages and differences, with the normalization factor of
$\frac{1}{\sqrt{2}}$. In the subsequent step, the average
coefficients are divided into two parts, containing level-II
high-pass and level-II low-pass coefficients. The high-pass
coefficients now represent differences of averaged data points
corresponding to a window size of two.  Thus higher level coefficients
represent lower frequency features.  Wavelets belonging to
the Daubechies family are designed such that the wavelet coefficients
are independent of polynomial trends in the data.
We have carried out a 10-level decomposition using
Daubechies-4 wavelets for isolating
fluctuations at different resolutions.
Daubechies-4 wavelet satisfies $ \int x \psi(x) dx=0$, in
addition to all other conditions. Because of this the wavelet
coefficients
capture fluctuations over and above the linear variations.  Simple
models of inflation predict that log $P(k)$ vs $\log (k)$ will have a
linear relation.  The choice of Daubechies-4 wavelets is the {\em minimal} (simplest) wavelet that allows us to
study variations about linear behavior in a window whose size
increases with the level of decomposition.

Furthermore, to study the significance of features located at different wavenumbers and at different resolutions, fluctuations associated with wavelet coefficients of different levels are added to the average behavior captured by the low-pass
coefficients in order to reconstruct a smoothened power spectrum.
Then a likelihood analysis with respect to the WMAP data is performed.

If the data set is of size $2^N$ a maximum of $N$ level decompositions
can be carried out. In the case of fewer data points, one needs to
supplement the data with additional points to carry out an $N$ level
decomposition. Due to the finite size of the filter coefficients, one
also encounters boundary artifacts due to circular or other forms of
extensions. In our case, for minimizing these boundary artifacts, we
have carefully padded the data with constants at both ends, which were
then removed after reconstruction.  We worked with a data set of 8192
points for log $P(k)$ vs $\log(k)$ recovered from WMAP-1 observations.
The data set was arranged to be equally spaced, as required by wavelet
programs.

\begin{figure}[h]
  \includegraphics[scale=0.8, angle=0]{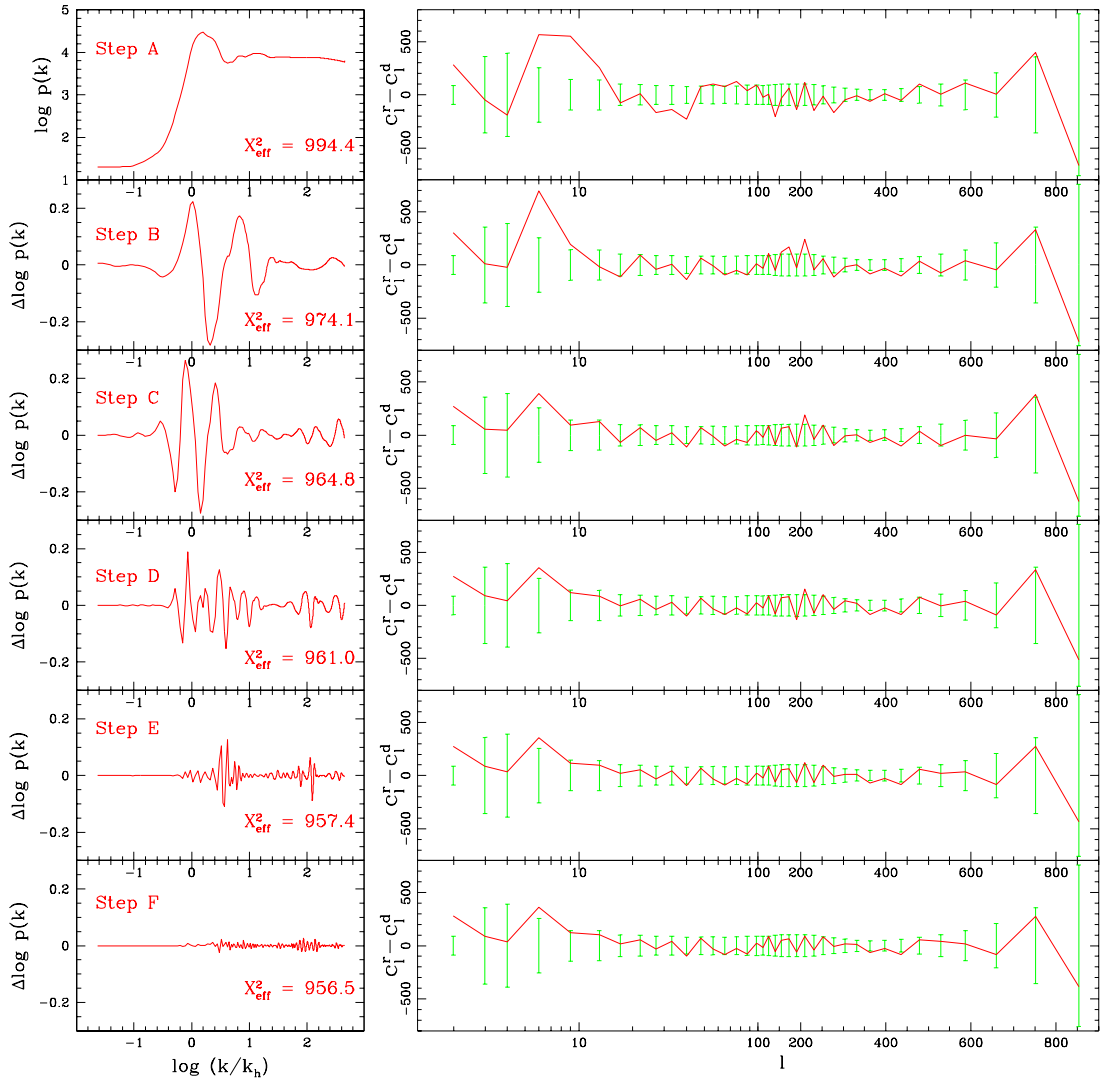}
  \caption{ The left panels show the DWT decomposition of the
  features of the primordial power spectrum. Step A shows the reconstructed
primordial power spectrum using the 10th level low-pass coefficients.
Steps $B, C, D, E$ and $F$ show the localized variations due to the
wavelet coefficients at
the 10th, 9th, 8th, 7th and the 6th levels respectively.  These
variations in the primordial power spectrum are most prominent
close to the horizon scale
and are significant only in the first few panels, corresponding
to `low frequency' variations.
The panels on the right compare the resultant
  angular power spectrum, $C_l^r$ with the binned WMAP-1 angular power spectrum
  data.
Going down from the top we progressively add
  features from different levels of wavelet coefficients (left)
to the primordial power spectrum and show the difference, $C_l^r-C_l^d$
and error bars for $C_l^d$ (right).
}
\label{wave}
\end{figure}

\begin{figure}[h]
  \includegraphics[scale=0.4, angle=0]{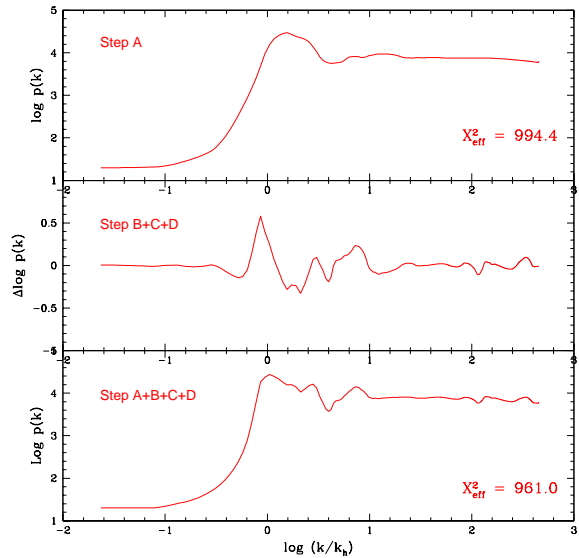}
  \caption{ The primordial power spectrum reconstructed with the
low-pass coefficients is shown in the top panel. Addition of features
of steps $B$, $C$ and $D$ is shown in the middle panel. The
combination of 
both these panels is shown in the bottom panel. Note the significant effect of
  the features on the likelihood.}
\label{feat}
\end{figure}

\begin{figure}[h]
  \includegraphics[scale=0.4, angle=0]{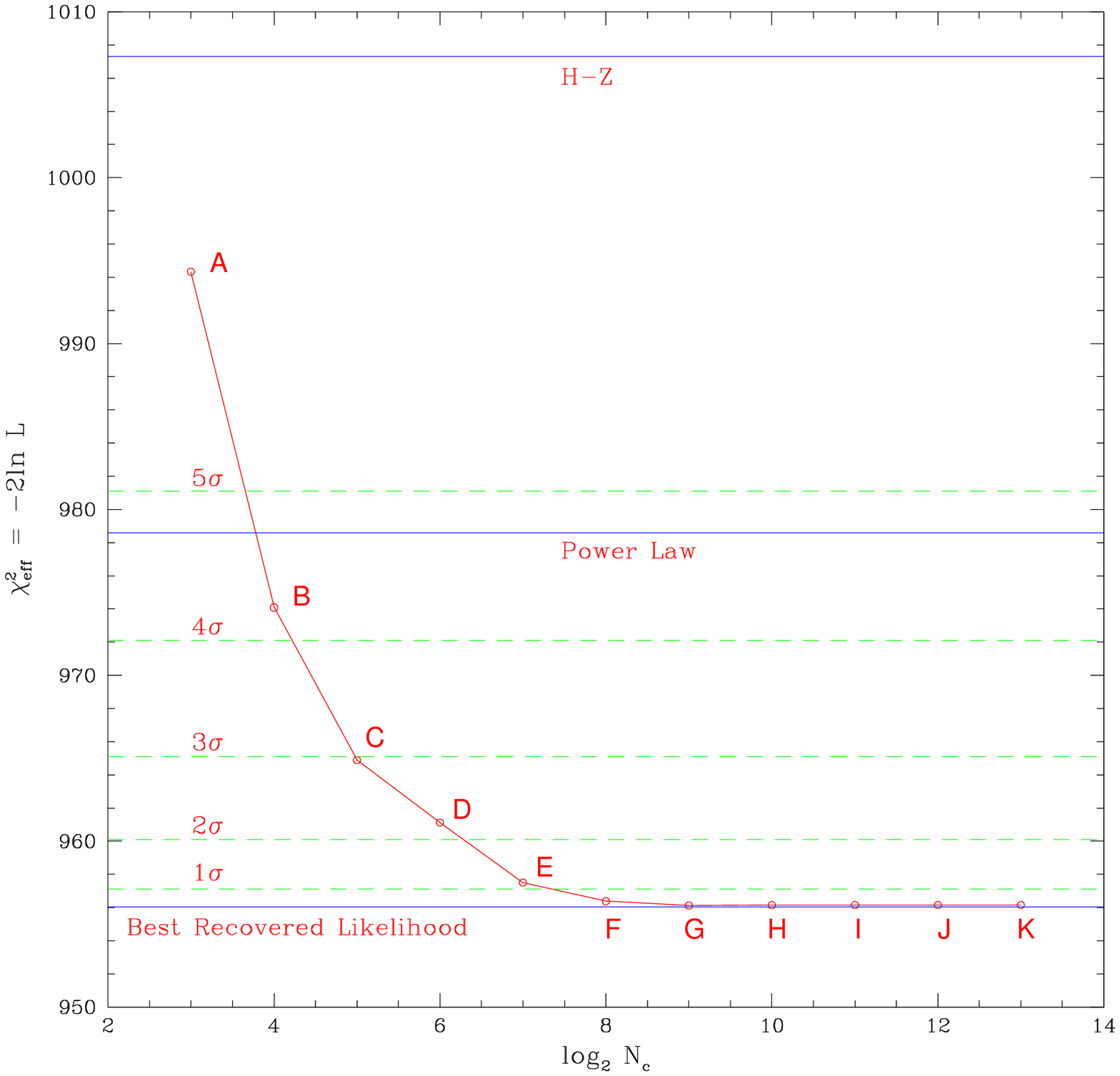}
  \caption{ The likelihood values improve as detailed features are
added to the primordial power spectrum reconstructed with low-pass
coefficients.  $A$ corresponds to the primordial power spectrum
reconstructed with low-pass coefficients only.  $B-K$ correspond to
the primordial power spectrum with the contribution of wavelet
coefficients of levels 10-1 progressively added to the coarse spectrum
$A$.  There is no significant improvement in the likelihood beyond
level 6 (step F).  Green-dashed lines mark the Gaussian equivalent
$\sigma$-levels of the likelihood relative to the best recovered
spectrum. Results are compared with the best likelihood given by Power
Law and Harrison-Zeldovich spectrum. $N_c$ is the number of
coefficients (low-pass and high-pass) used to define the
primordial spectrum at each step.}
\label{like}
\end{figure}

\section{Features of the primordial power spectrum}

One of the most challenging questions of the modern cosmology is to
find an inflationary scenario satisfying all the cosmological
observations. The shape of the primordial power spectrum has the key
role in this investigation. In our previous paper \cite{prd04} we have
used Richardson-Lucy deconvolution algorithm to find the shape of the
primordial power spectrum using the cosmic microwave background data.

By using DWT, we decompose the recovered primordial power spectrum
locally at different resolutions.  We then calculate the angular power
spectrum after including the variations in the primordial power
spectrum at different resolutions. We subsequently compute the likelihood
of the primordial power spectrum at each of these stages so as to
quantify the effect of different features on the recovered angular
power spectrum.  We use WMAP-1 likelihood code available at the LAMBDA
site ~\footnote{Legacy Archive for Microwave Background Data Analysis
(LAMBDA) webpage: http://lambda.gsfc.nasa.gov/} and quote our likelihood $\cal L$ in terms of
$\chi^{2}_{\rm eff}=-2\ln {\cal L}$.

The original primordial power spectrum which we use in this paper (see figure
\ref {main}) is the final recovered primordial power spectrum obtained in 
Ref.~\cite{prd04} for a flat $\Lambda$CDM cosmological model
with $\tau$=0.0, $h$=0.71, $\Omega_b h^2$=0.0224 and
$\Omega_{\Lambda}$=0.73. The resultant $C_l$ spectrum
using this recovered spectrum has a likelihood far better than a scale
invariant or a best fit power law spectrum. In this 
paper, we investigate how
the features of this spectrum contribute to improving the likelihood.

 The blue dotted line in figure \ref {main} represents the recovered primordial spectrum by using WMAP-3 data which gives $\Delta \chi^2_{eff} =-15.93$ (by using WMAP-3 likelihood code) with respect to the best power law primordial spectrum in the whole cosmological parameter space. In this paper we are not using this recovered spectrum for our wavelet analysis and it has been presented here just to show the close similarity of the features of the recovered spectrum from WMAP-1 and WMAP-3 data.
 
First, we smooth the spectrum using DWT, and then we systematically include
features at different resolutions and calculate the
likelihood with respect to the WMAP data.

In figure \ref {wave}, left panel (step A), we show the coarsest
behavior of the data reconstructed using only the 10th level low-pass 
coefficients. The right 
panel shows the resultant $C_l^r$ compared with the observed binned
data $C_l^d$, and its error bars. The likelihood of the $C_l^r$ at this
stage corresponds to $\chi^2_{\rm eff}=994.4$ which is better than the $\chi^2_{\rm eff}$ for the H-Z
spectrum. 
In the right panel (step B) we see that if we include 
`low frequency' features as captured by the 10th level high-pass coefficients 
shown in the left panel (step B) to the previous spectrum at step A, 
the resultant $C_l^r$ will be closer to the observed data.  Hence we expect the
likelihood to be improved. In fact this improves the likelihood significantly 
and the $\chi^2_{\rm eff}=974.1$ at this stage (the reconstructed spectrum is 
the green-dashed line in figure~\ref {main}). This $\chi^2_{\rm eff}$ is better than the 
best fit power law spectrum with the $\chi^2_{\rm eff}=978.6$.

As we
progressively add back more features at `higher frequencies' to the
spectrum the likelihood improves significantly.
The 9th, 8th, 7th and 6th level wavelet coefficients 
are shown in the left column of figure~\ref {wave} (steps $C, D, E$ and $F$).
The plots in the right panel of figure~\ref{wave}
show how we get
closer to the observed data within the error bars
as we include more wavelet coefficients.
 The likelihoods at these steps correspond to 
are $\chi^2_{\rm eff}=964.8$ at step $C$, $\chi^2_{\rm eff}=961.0$ at step $D$, $\chi^2_{\rm eff}=957.4$ at step $E$, and
$\chi^2_{\rm eff}=956.5$ at step $F$.

In figure~\ref {feat} we compare the coarsest spectrum of step $A$
and the spectrum after adding the features of step $B$, $C$ and $D$.  The
local features which are responsible for the significant improvement of the
likelihood are clearly seen.
We note that most of these features are localized around the horizon scale. In figure \ref {like}, we 
see that after few stages of adding the more detailed features to the
spectrum, the likelihood does not improve anymore and agreement with
the observed
data is not sensitive to these features.

\section{Conclusion}
 
This paper presents a detailed analysis of the recovered primordial
power spectrum for a flat $\Lambda$CDM cosmological model, given in
Ref.~\cite{prd04}. The recovered spectrum has a likelihood far better
than a simple scale invariant or a scale free spectra. In Ref.~\cite{bump05}, similar features for the primordial spectrum have been
detected by using a completely different method and the significance
of the features have also been evaluated. In this paper we use
Discrete Wavelet Transform to decompose the features of the
spectrum to quantify and understand their role in improving the
likelihood. In addition to the infra-red cut-off around the horizon, which was
proposed by many groups to explain the lack of power in very low
multipoles of the observed angular power spectrum, we show that the
features around the horizon are playing a crucial role in improving
the likelihood. In fact, the effect of these features on improving the
likelihood, are very significant (figure ~\ref {feat}). We find that
these strong features are localized around the horizon. Work on WMAP 3
years data is in progress and will be reported in a forthcoming
publication~\cite{armarun06}. The work in progress implies refinement of methodology includes directly employing DWT
right at the stage of smoothing the raw recovered spectrum and
dispensing with the need to remove a `template' spectra as in
Ref.~\cite{prd04}.


\section{Acknowledgment}     
We thank S. Hemachander and S. Nanavathy for their contribution in the
early stages of this work. We acknowledge the use of the Legacy Archive for Microwave Background Data Analysis (LAMBDA).  Support for LAMBDA is provided by the NASA Office of Space Science.

\newpage
\newpage


\begin{thebibliography}{99}
\bibitem{notearmarun06} We can dispense with the step of removing a template employed in Ref.~\cite{prd04} to remove the artifacts at low and high wavenumber with a small change in our algorithm ~\cite{armarun06}.
\end{thebibliography}
\end{document}